# FACTS, MYTHS AND FIGHTS ABOUT THE KLJN CLASSICAL PHYSICAL KEY EXCHANGER

LASZLO B. KISH

*Department of Electrical and Computer Engineering, Texas A&M University, College Station, Texas 77843-2128, USA*

*laszlo.kish@ece.tamu.edu*

DEREK ABBOTT

*School of Electrical and Electronic Engineering, University of Adelaide, Adelaide, South Australia, Australia*

*dabbott@eleceng.adelaide.edu.au*

CLAES-GÖRAN GRANQVIST

*Department of Engineering Sciences, The Ångström Laboratory, Uppsala University, P.O. Box 534, SE-75121 Uppsala, Sweden*

*claes-goran.granqvist@angstrom.uu.se*

HE WEN

*College of Electrical and Information Engineering, Hunan University, Changsha, China*

*he_wen82@126.com*

This paper deals with the Kirchhoff-law–Johnson-noise (KLJN) classical statistical physical key exchange method and surveys criticism—often stemming from a lack of understanding of its underlying premises or from other errors—and our related responses against these, often unphysical, claims. Some of the attacks are valid, however, an extended KLJN system remains protected against all of them, implying that its unconditional security is not impacted.

The Kirchhoff-law–Johnson-noise (KLJN) classical statistical physical key exchange method [1–12] offers unconditional (information theoretic) security without quantum physics by utilizing the laws of classical statistical physics. This paper outlines a selection of our published responses to various criticisms of the security of the KLJN scheme. Some of the attacks were valid, such as the (Bergou-) Scheuer-Yariv resistance attack [11] and Hao's temperature attack [12] however by investing sufficient resources the leaked information can be made arbitrarily small and hence its unconditional security is not impacted [1]. Furthermore, when the resources are limited, the extended KLJN system (that includes public comparison of data measured by Alice and Bob and the related rejection of bits when their security is compromised beyond a chosen threshold) is naturally protected against all of them. Recent important development is that all wire resistance based attacks can be eliminated by a simple trick, by introducing another proper non-ideality feature see [13]. Several other attacks originated from the lack of understanding of the KLJN scheme or that of the definition of unconditional security.

*The conclusion of these debates is that the unconditional security of the existing KLJN schemes stands firm.*

The most important aspects of the talk have been addressed in publications (e.g., [1–8]) and are briefly presented below. Three essential points concerning the KLJN scheme are as follows:

**1.** Unconditional security of the ideal KLJN scheme at passive (i.e. listening) attacks is guaranteed by the *Second Law of Thermodynamics*. Thus this type of attack is as hopeless as efforts to make a perpetual motion machine (of the second kind). Thus the KLJN scheme has perfect security in this case [1]. Moreover, this situation of perfect security holds even if the non-ideal feature of a non-zero wire resistance is introduced, see [13].

**2.** In general, unconditional security of the non-ideal KLJN scheme at passive attacks is guaranteed [1] by the *continuity of functions in stable classical physical systems*, which means that a perfect security level cannot be reached at non-zero temperature, but it can be approached asymptotically. This is the exact definition of unconditional security.

**3.** Unconditional security of the ideal and non-ideal KLJN system at active (i.e., invasive) attacks is guaranteed by a current–voltage comparison of Alice's and Bob's data—using conventional notation—via a publicly authenticated channel. This means that a perfect security level cannot be reached at non-zero speed, but it can be approached asymptotically [1]. This situation again implies unconditional security.

4. Defense against hacking attack must involve a full statistical and frequency analysis of voltages are currents by Alice and Bob moreover the random/arbitrary monitoring of the information channel (wire) for the integrity of its system parameters [13].

Finally, one typical misunderstanding is clarified [2]: Eve can have infinite measurement speed and accuracy, but still her information is strongly limited by basic laws of information theory and signal processing! The reason is that noise in the KLJN scheme is band-limited, as apparent from its power density spectrum $S(f)$ illustrated in Figure 1.

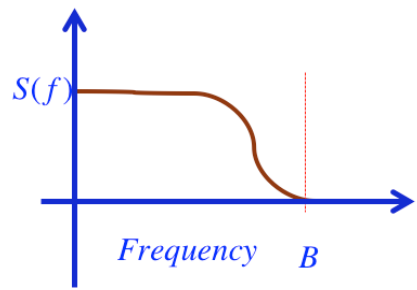

**Fig. 1.** Illustration of the band-limited spectrum $S(f)$ of noise in the KLJN system. $B$ denotes bandwidth.

According to the Nyquist–Shannon sampling theorem, during a bit exchange period $\tau$ a measurement can extract only $n \leq 2B\tau$ independent samples of the measured noise, where $B$ is bandwidth, even if Eve has "perfect" measurement instruments with infinite bandwidth and accuracy. Alice and Bob have full control over $n$ because they set the bandwidth and the duration of the single-bit exchange. Eve's only ways to extract information are to make statistics of the noise under invasive (active) attacks or to exploit non-ideal features based on second-(or higher)-order effects. In both cases, Alice and Bob can ascertain that Eve's sample number remains insufficient [2], which means that eavesdropping does not occur. Furthermore, Alice and Bob can limit Eve's information still more by discarding high-risk bits that provide information above a publicly agreed threshold [1].

Finally, some additional publications to refer to; math proofs by Gingl and Mingesz that only Gaussian noise can provide security [14,15]; statistical analysis of the enhanced KLJN system by Smulko [16]; and analysis of the multiple flaws of a recently proposed invalid attack claim including and exact proof that the assumption of waves in the KLJN regime violates several basic laws of physics [17,18].